# On the use of financial analysis tools for the study of $D_{st}$ time series in the frame of complex systems

Stelios M. Potirakis [a, *], Pavlos I. Zitis [b], Georgios Balasis [c]
and Konstantinos Eftaxias [b]

*a. Department of Electronics Engineering, Piraeus University of Applied Sciences (TEI of Piraeus), 250 Thivon and P. Ralli, Aigalao, Athens GR-12244, Greece, spoti@teipir.gr*

*b. Department of Physics, Section of Solid State Physics, University of Athens, Panepistimiopolis, GR-15784, Zografos, Athens, Greece, ceftax@phys.uoa.gr*

*c. Institute for Astronomy, Astrophysics, Space Applications and Remote Sensing, National Observatory of Athens, Metaxa and Vasileos Pavlou, Penteli, 15236 Athens, Greece, gbalasis@noa.gr*

**Abstract**

Technical analysis is considered the oldest, currently omnipresent, method for financial markets analysis, which uses past prices aiming at the possible short-term forecast of future prices. In the frame of complex systems, methods used to quantitatively analyze specific dynamic phenomena are often used to analyze phenomena from other disciplines on the grounds that are governed by similar dynamics. An interesting task is the forecast of a magnetic storm. The hourly $D_{st}$ is used as a global index for the monitoring of Earth's magnetosphere, which could be either in quiet (normal) or in magnetic storm (pathological) state. This work is the first attempt to apply technical analysis tools on $D_{st}$ time series, aiming at the identification of indications which could be used for the study of the temporal evolution of Earth's magnetosphere state. We focus on the analysis of $D_{st}$ time series around the occurrence of magnetic storms, discussing the possible use of the resulting information in the frame of multidisciplinary efforts towards extreme events forecasting. We employ the following financial analysis tools: simple moving average ( *SMA* ), Bollinger bands, and relative strength index ( *RSI* ). Using these tools, we formulate a methodology based on all indications that could be revealed in order to infer the onset, duration and recovery phase of a magnetic storm, focusing on the temporal sequence they occur. The applicability of the proposed methodology is examined on characteristic cases of magnetic storms with encouraging results for space weather forecasting.

## 1. Introduction

The relatively new field of complex systems increasingly gains the interest of scientists working on disciplines ranging from physics and engineering to economics, biosciences and social sciences, e.g., [1-8]. The unique characteristic of complex systems is that they may have certain quantitative features that are intriguingly similar, while their dynamics are governed by a set of universal principles [9, 10]. Thus, complex systems from different disciplines are often analyzed within similar mathematical frameworks.

There is an apparent paradox in the above mentioned suggestion. How is it possible for a concept as multifaceted as complexity to serve as a unifying direction? When one considers a phenomenon that is "complex'" refers to a system whose phenomenological laws, which describe the global behavior of the system, are not necessarily directly related to the "microscopic" laws that regulate the evolution of its elementary parts [11]. This is a basic reason for our interest in complexity [11-15]. There is a common factor in these seemingly diverse phenomena. The complex systems adopt a pattern of behavior almost completely determined by the collective effects. They exhibit remarkable properties of self-organization and emergence of coherent structure over many scales. The main feature of collective behavior is that an individual unit's action is dominated by the influence of its neighbors; the unit behaves differently from the way it would behave on its own, so that, all units simultaneously alter their behavior to a common pattern. Thus, new features emerge as we move from one scale to another and the science of complexity is about revealing the principles that govern the ways in which new properties appear [11].

Two of the most vivid and richest examples of the dynamics of a complex system at work is that of economic systems (financial markets) [14, 16-24] and Earth's magnetoshere [25-35]. Their richness in interactions renders them characteristic examples of complex dynamics. In this paper, we focus on the identification of indications which could be used for the study of the temporal evolution of Earth's magnetosphere state and consequently for the forecast of magnetic storms (MSs). Magnetic storms occur when the accumulated input power from the solar wind to magnetosphere exceeds a certain threshold. Notice, MSs are a main element of space weather: they have severe impacts on both space-borne and ground-based technological systems [36]. Thus, the prediction of a forthcoming intense MS is a major task. Magnetic storm intensity is usually represented by an average of the geomagnetic perturbations measured at four mid-latitude magnetic observatories, known as the $D_{st}$ index [37].

This work aims at enhancing the suggestion that transferring ideas, methods and insights from investigations in hitherto disparate areas, namely, economic and geophysical systems, will cross-fertilize and lead to important new results concerning the dynamics of the corresponding extreme events, i.e., dynamics of economic crises and magnetic storms. A question effortlessly arising is whether the aforementioned notion is groundless or not. A number of analysis methods have been mutually used to study the dynamics of financial markets, earthquakes, and magnetosphere, e.g., [17, 38-50, 32-35]. Several authors have suggested that earthquake dynamics and the dynamics of economic (financial) systems can be analyzed within similar mathematical frameworks, e.g., [10, 14, 51-54, and references therein]. On the other hand, authors have also suggested that earthquake dynamics and magnetic storms dynamics can be analyzed within similar mathematical frameworks, as well e.g., [33-35]. Thus, the question whether these two complex systems, namely, financial crises and magnetic storms, can be analyzed within the same mathematical framework seems to be justified. This conclusion is specifically enhanced by the fact that the signature of Discrete Scale Invariance (DSI) [49] characterizes the earthquakes and financial crises [49], and magnetic storm [34], as well. It is important to stress the practical consequence of the presence of the corresponding log-periodic structures. For prediction purposes, it is much more constrained and thus reliable to fit a part of an oscillating data than a simple power law which can be quite general especially in the presence of noise [49].

Technical analysis, primarily employed for the empirical analysis of economical time series, is considered the oldest method for investment analysis with origins dating perhaps before the 1800s [55]. Charles Dow who developed the famous Dow Theory, which was later refined by S. A. Nelson, W. P. Hamilton and R. Rhea in the early 20$^{th}$ century, is considered the pioneer of modern technical analysis [56]. Nowadays, technical analysis is omnipresent in financial markets analysis. Taylor and Allen [57] as well as Menkhoff [58], nearly two decades later, analyzed data of 692 fund managers in five countries, including the USA, concluding that at least 90% or 87%, respectively, of the involved respondents pay enough attention to technical analysis in order to take investment decisions primarily on short term investments. Note that technical analysis has also been used in other disciplines too such as medicine, e.g., [59, 60], data communications, e.g., [61, 62], textile engineering, e.g., [63], wireless sensor networks, e.g., [64] and aviation [65]. Technical analysis uses past prices having as target the possible identification of future prices. The efficiency of technical analysis in markets which are characterized by long-term memory, as determined by Hurst exponent, has been recently studied in fifteen of the largest global equity markets [55].

In this paper, we investigate the use of some widely employed tools of technical analysis for the study of the temporal evolution of $D_{st}$ time series. Specifically, we study here whether the simple moving average ($SMA$), the Bollinger bands, and the relative strength index ($RSI$) can discriminate in the $D_{st}$ time series the transition from pre-storm (quiet) geomagnetic activity to the generation of a magnetic storm, through the identification of specific indications in particular temporal sequence. Note that, to the best of our knowledge, this is the first time that such a study appears in the literature. Such a study: (i) could be utilized for space weather forecasting purposes, especially if it will be combined with other studies (e.g., of log-periodic corrections) and (ii) may offer ideas in order to provide a physical meaning in the above mentioned empirical tools of technical analysis.

The remaining of this contribution is organized as follows: The necessary background information on the used technical analysis tools, as well as information about the analyzed time series is provided in Sec. 2. In Sec. 3 we describe the behavior of the considered technical analysis tools when applied on $D_{st}$ time series during quiet magnetospheric conditions as well as during the evolution of a magnetic storm, while a methodology for the use of the specific tools in the study of magnetic storms is introduced as a step-by-step procedure. The analysis of different $D_{st}$ time series including magnetic storms is presented and the obtained results are discussed in Sec. 4. Finally, in Sec. 5, the presented $D_{st}$ analysis method and obtained results are discussed, while the conclusions are summarized.

## 2. Data and Analysis Methods

In the following we provide a brief introduction to the main information and equations related to the analysis methods which are used in this paper, as well as a description of the Earth's magnetosphere observables ($D_{st}$ time series) on which these methods are applied on.

### 2.1. Methods

The main aspects of the simple moving average ($SMA$), Bollinger bands, and relative strength index ($RSI$) methods which are widely used in the empirical analysis economic time series known as "technical analysis" are briefly presented in the following. The introduction to these methods is presented in terms of the information which is extracted by each one of them during the analysis of stock market time series, while all necessary mathematical formulas for their application are also given.

#### 2.1.1. Moving Average

The moving average is one of the most popular and widely used tools of technical analysis. It is characterized as an automated trendline and it is considered an objective and reliable tool for stock market analysis [66, 67]. The information provided by moving average is threefold [68]: (i) it determines the direction of the trend, (ii) it confirms the change of the trend and (iii) it smoothes the extraneous data which are often misguiding. As it is expected, the moving average responds to price changes with lag, since it is calculated using past prices. The longer the length of the calculation period, the slower it responds to changes [69]. A wide variety of moving averages are used in stock market time series analysis. In this paper we will refer to the simple moving average ($SMA$) which is defined by the following formula [56]:

$$SMA_k = \frac{1}{n} \sum_{i=k-n+1}^{k} C_i, \qquad (1)$$

where $SMA_k$ is the simple moving average at period $k$, $C_i$ is the closing price for the period $i$, $n$ is the total number of periods to be included in the moving mean calculation and $k$ is the number of the period being studied within the total number of periods in the database.

The most popular interpretations of the moving average results are the following: (i) *upward (or downward) crossing of the moving average curve by the price curve is a sign of upward (or downward) trend* [70]; (ii) *upward (or downward) slope of moving average curve means upward (or downward) trend of the price; also if the slope is steep, the trend is expected to be strong, while, conversely, if the slope is gentle, the trend is expected to be weak* [71]; (iii) *the moving average often acts as "support" and "resistance" level of the price trend* [72], i.e., if the price values and the moving average are plotted on the same chart, and the chart shows that the price value, within a time period, never seems to be able to rise above a value A or fall below a value B of the moving average, then the resistance level is the value A and the support level is the value B, respectively; (iv) *the prices have the tendency to return to the value of moving average* [72].

### 2.1.2. Bollinger Bands

The Bollinger bands were developed by John Bollinger in the 1980s and are a tool of technical analysis that belongs to a wider category of price analysis methods called "trading bands". The trading bands in general consist of curves lying at a distance above and below a measure of central tendency [73, 74]. In the specific case of Bollinger bands the measure of central tendency is the simple moving average and the distance of the curves from the moving average depends on the local standard deviation of prices. The standard deviation measures the volatility of prices, since it is a statistical measure that indicates how far prices range from average [75]. Due to the way Bollinger bands are constructed, their distance regulates itself and adapts to the volatility of market prices. When volatility is high, the bands widen, while when it is small the bands narrow approaching the moving average [76]. The Bollinger bands consist of three curves: the Middle Band ($MiddleB$), the Upper Band ($UpperB$), and the Lower Band ($LowerB$), calculated by the following formulas, respectively:

$$MiddleB = SMA, \qquad (2)$$

$$UpperB = SMA + K \cdot \sigma_n, \qquad (3)$$

$$LowerB = SMA - K \cdot \sigma_n, \qquad (4)$$

where $n$ is the calculation period of $SMA$, $\sigma_n = \sqrt{1/n \sum_{i=k-n}^{k}(x_i - SMA)^2}$ is the standard deviation of the same period and $K$ is a parameter that determines the distance of the bands from the moving average.

The originally proposed pairs of $n$ and $K$ are [73]: $(n=10, K=1.9)$, $(n=20, K=2)$, and $(n=50, K=2.1)$. As it is expected, these parameters were selected so that the majority of the prices fall within the bands. It has been argued that, in most markets, about 88% to 89% of the prices are within the bands for $n=20$ and $K=2$ [73]. This has been further supported by the work of Liu et al. [76] who studied stock market indices (DOW, NASDAQ and S&P500) from 1991 to 2005 concluding that 94% of the daily closing prices were within the Bollinger bands. Moreover, Xu and Yang [77] reached a similar conclusion after studying the indices SPY, QQQ and DIA for the period 2008 to 2011, namely during the economic crises of 2008, and found that more than 95% of the prices lie within Bollinger bands [77].

The most popular interpretations of Bollinger bands results are the following [70]: (i) *after the narrowing of the bands, the prices usually have a sharp change;* (ii) *when prices move outside the bands, this implies that the current trend is going to continue;* (iii) *when local maximum (or local minimum) which is outside of bands is followed by local maximum (or local minimum) inside the bands, this is an indication of an upcoming reversing of the current trend.;* (iv) *for a price change which starts at one of the bands it is expected that this will continue changing, maintaining its trend, until it covers all the way to the other band.*

Finally, we should mention that originally the Bollinger bands method was exclusively used in finance time series analysis, but over time, this method has been proven to find application in various fields such as textile engineering, data networks, wireless sensor network and aviation, e.g., [62-65].

### 2.1.3. Relative Strength Index (RSI)

Relative Strength Index ($RSI$) was developed by W. J. Wilder [78]. It is one of the best known momentum indicators and measures the speed and the magnitude of the direction of price movements [79, 77]. The $RSI$ takes values in the interval [0, 100]; generally, values of $RSI$ over 70 suggest probable "overbought" situation, while values below 30 indicate a probable "oversold" situation. Overbought (or oversold) is a term used to describe the situation following a rise (or decline) in share prices, in which some investors believe that prices have risen (or fallen) exceedingly, i.e., the share is overvalued (or undervalued) relative to fundamentals [80]. The specific threshold values (70 and 30) are not absolute; other threshold values also appear in the literature [72]. The $RSI$ value is calculated using the following equation [78]:

$$RSI = 100 - \left[\frac{100}{1+RS}\right], \qquad (6)$$

where $RS = \dfrac{\text{moving average of } n \text{ day's closes UP}}{\text{moving average of } n \text{ day's closes DOWN}}$, the originally suggested calculation period for the moving average is $n=14$ [78]. However, other values for the calculation period, $n$, have also been considered later, with most popular being: 5, 7, 8, 10 and 20 days [56]. Apparently, the smaller the calculation period, $n$, the more sensitive to price changes is the $RSI$, while in the opposite case the index is less sensitive [56].

The following interpretations of the $RSI$ analysis results have been proposed [78]: (i) *the RSI often reaches a zero slope (local maximum) over 70 or a zero slope (local minimum) below 30, prior to a local maximum or local minimum of the prices, respectively, providing an indication of a possible trend reversal;* (ii) *the chart of RSI shows often formations, such as "head" and "shoulders"* (when three peaks appear successively and the second is higher than the other two, then the first and third are called shoulders and the second is called head) *or "triangles"* (a triangle occurs as the range between peaks and troughs narrows)*, that indicate a possible change of trend, formations sometimes not evident in the price chart;* (iii) *"failure swings" above the level of 70 (or below the level of 30) are very strong evidence of a trend reversal.* The term failure swing describes a local maximum (or local minimum) of the $RSI$ curve above 70 (or below 30) that is followed by a second local maximum (or local minimum) below 70 (or above 30)*;* (iv) *the RSI shows several times clearer "resistance" and "support" levels of the chart of prices*, i.e., if the $RSI$ value is plotted on a chart, and the chart shows that the $RSI$ value, within a time period, never seems to be able to rise above a value A or fall below a value B, then the resistance level is the value A and the support level is the value B [80]*;* (v) *when "divergence" between the prices and the RSI values are observed,* namely new high (or low) that is not verified by a new high (or low) of the $RSI$ values*, the prices tend to follow the direction of motion of the RSI values.*

### 2.2. Data

Magnetic storms are the most prominent global phenomenon of geospace dynamics, interlinking the solar wind, magnetosphere, ionosphere, atmosphere and, occasionally, the Earth's surface [36, 81, 82]. Magnetic storms produce a number of distinct physical effects in the near-Earth space environment: acceleration of charged particles in space, intensification of electric currents in space and on the ground, impressive aurora displays and global magnetic disturbances on the Earth's surface [83]. The latter serve as the basis for storm monitoring via the hourly $D_{st}$ index, which is computed from an average over four mid-latitude magnetic observatories [37]; $D_{st}$ is considered a global index for the monitoring of Earth's magnetosphere, The size of a geomagnetic storm is classified as moderate (-100 nT < minimum of $D_{st}$ < -50 nT), intense (-250 nT < minimum $D_{st}$ < -100 nT) or super-storm (minimum of $D_{st}$ < -250 nT).

The here proposed methodology was extracted after the analysis of more than 20 magnetic storms of all classes occurred from 1958 to 2015; their time of occurrence and magnitudes are shown in the second and third column of Table 1, respectively. An illustrative example of application is given in parallel to the presentation of the proposed methodology (cf. Sec. 3.2) using a super-storm which occurred on 31/3/2001 with peak $D_{st}$ value of -387 nT (line no. 13 of Table 1). Moreover, we present in detail (in Sec. 4) the analysis of four characteristic cases of magnetic storms which occurred on 26/05/1967, 22/10/1999, 08/11/2004 and 25/10/2011 with peak $D_{st}$ values of -377 nT, -237 nT, -374 nT and -147 nT, shown in lines no. 8, 10, 18 and 23 of Table 1, respectively. Informative plots summarizing the analysis results for the full set of analyzed magnetic storms listed in Table 1 are provided as on-line supplementary material.

## 3. A Proposed Magnetic Storm Analysis Based on Stock Market Tools

Technical analysis is the process aiming at the possible identification of future prices by analyzing past price data [70], using a set of empirical stock market analysis tools. Crucial factor towards achieving this goal is to identify the direction (upward, downward or "sideways" –horizontal–), duration and strength of the price trend [79]. In order to identify those characteristics of trend, the use and combined interpretation of different tools of technical analysis is required [73, 79]. The application of these tools is flexible and can be adapted to any time duration (from a few minutes up to months) [56]; however, the obtained results are considered more reliable in the case of short-term (in the scale of a few hours to a few days) analyses, compared to the ones obtained for long-term (in the scale of a many months to a few years), ones [75]. Based on this fact and given that the duration of a typical magnetic storm ranges in the order of a few hours up to a few days, while the available data are hourly $D_{st}$ values, a short term prediction of the time evolution of the phenomenon seems feasible.

In this paper, we attempt to apply a combination of three technical analysis tools, on hourly $D_{st}$ data variations. Specifically:

(i) The $SMA$, which has been calculated for $n = 20$ hours (cf. Eq. (1) ), is used to identify the trend of $D_{st}$ values and offers, albeit with some lag, quite reliable signs of upcoming changes in $D_{st}$ values. Note that, "if the slope is sharp, the trend is strong, and if the slope is shallow, the trend is weak; a flat or choppy *SMA* indicates a range-bound market" [71].

(ii) The Bollinger bands, calculated for an *SMA* of $n = 20$ hours and $K = 2$ (cf. Eqs. (2) – (4)), are used in order to provide a depiction of the range of variation of the $D_{st}$ values. Observing the width of the bands and the movement of the $D_{st}$ curve within or outside the bands, we draw conclusions for the volatility of $D_{st}$ values as well as for the duration that this would have. When combined with other indicators, such as the *RSI*, the Bollinger Bands become quite powerful, since *RSI* is "an excellent indicator with respect to overbought and oversold conditions" [67].

(iii) The *RSI* indicator, which has been calculated for $n = 20$ hours and is a momentum indicator, is used to measure the speed and strength of the trend of $D_{st}$ values. When the *RSI* curve falls below 30, this indicates that $D_{st}$ values have been reduced considerably in a short period of time.

Through the application of these stock market tools on $D_{st}$ time series, our aim is to identify specific indications which could be used for the study of the temporal evolution of the state of Earth's magnetosphere and, accordingly, formulate a methodology that could be employed in order to infer the onset, duration and recovery phase of a magnetic storm. The here proposed methodology was extracted after the analysis of 24 magnetic storms of all classes occurred from 1958 to 2015, shown in Table 1.

In the following Sec. 3.1 and Sec.3.2, we describe the behavior of the considered technical analysis tools during quiet magnetospheric conditions as well as during the evolution of a magnetic storm, respectively, providing thus a method of application of these tools for the study of the temporal evolution of $D_{st}$ time series. Especially in the second case, we provide a detailed description of all the indications that could be employed in order to infer the onset, duration and recovery phase of a magnetic storm, focusing on the temporal sequence in which they occur. The application of the proposed methodology on characteristic cases of magnetic storms is demonstrated in Sec. 4.

### 3.1. Quiet Magnetospheric Conditions

In periods that normal (quiet) magnetospheric conditions prevail, cf. Fig. 1, we observe that all the applied technical analysis methods suggest that no significant downward trend of $D_{st}$ values, which could possibly evolve to a magnetic storm, is expected. In other words, the values of $D_{st}$ (blue/solid curve in Fig. 1.a) are expected to follow a low amplitude variation around its mean value (expressed by the value of *SMA* which is portrayed as the green/dashed curve in Fig. 1.a), while the *SMA* curve evolves with nearly horizontal slope. In terms of the Bollinger bands (shown by the two red/dotted curves in Fig. 1.a), the same situation is reflected to that these bands are evolving close to each other, nearly horizontally, indicating low volatility of $D_{st}$ values; in most cases $D_{st}$ values vary within the Bollinger bands, however in some cases short $D_{st}$ curve segments may be found to move slightly outside one of the bands, as shown in Fig. 1.a (close to 25/10/2007 17:00). At the same time, the *RSI* indicator (shown in Fig. 1.b) mainly varies between the thresholds 30 and 70 (grey horizontal dotted and dashed line, respectively, in Fig. 1.b). However, several times "support" or "resistance" may be found at the threshold levels 30 or 70 respectively, i.e., the *RSI* value may be shortly found on (or slightly outside) the lower or the higher threshold level, respectively, to return inside immediately after. Note that a short movement of the *RSI* value in the oversold area (below 30) by itself, i.e., without being part of the sequence of indications described in Sec. 3.2 does not imply an impeding magnetic storm. The described behavior of the considered technical analysis methods can easily be verified in Fig. 1.

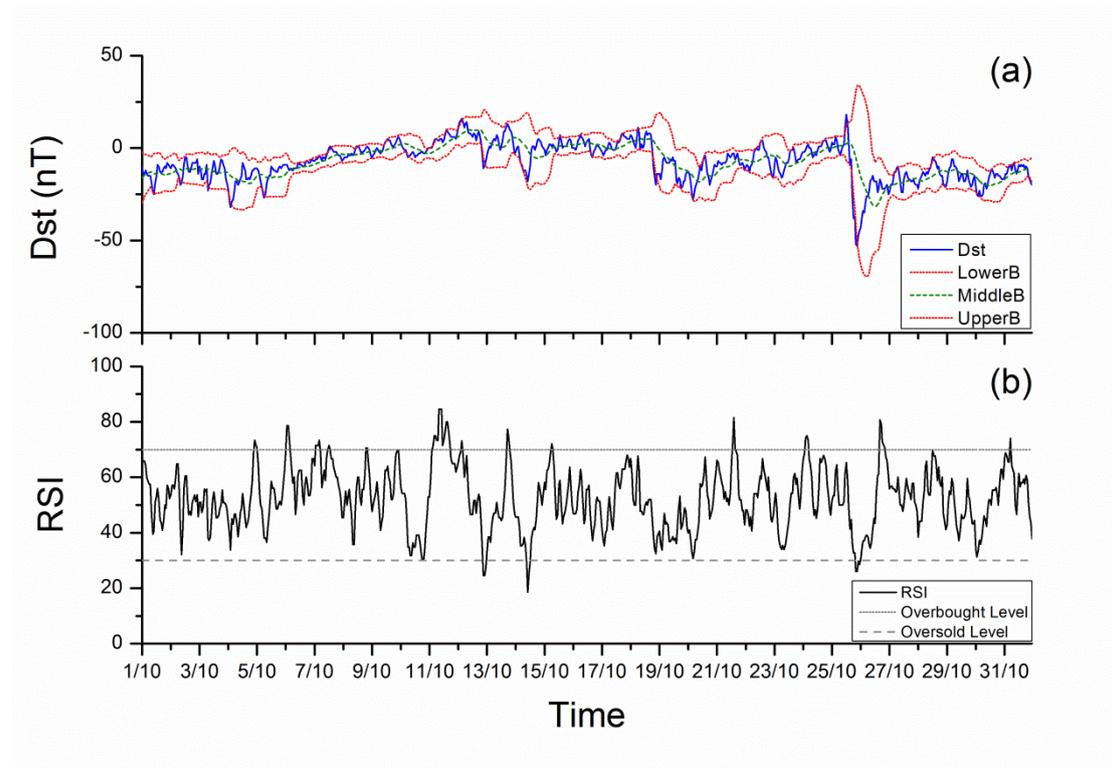

**Fig. 1**. The behavior of *SMA* and Bollinger bands (a), as well as *RSI* indicator (b) in periods that normal (quiet) magnetospheric conditions prevail. The analyses results for a period of one month from 01/10/2001 to 31/10/2001 are presented as an example.

## 3.2. Magnetic Storms Evolution in Terms of Financial Analysis Tools

After analyzing the $D_{st}$ time series corresponding to all the magnetic storms included in Table 1, we considered that it is appropriate to classify the indications, which resulted from the application of the employed technical analysis tools and could be used in order to infer the onset, duration and recovery of a magnetic storm, into main and secondary. The ones characterized as main are those which have been found to occur for all the examined cases of magnetic storms, whereas the indications characterized as secondary should be considered as indications supporting the main ones, without however being expected to be found in all cases.

On one hand, the main indications for the analysis of the evolution of a magnetic storm are: (M.i) the narrowing of the Bollinger bands; (M.ii) the downward crossing of the *SMA* curve by the $D_{st}$ curve; (M.iii) the steep downward slope of the *RSI* indicator; (M.iv) the downward crossing of the lower Bollinger band by the $D_{st}$ curve; (M.v) the retreat of the *RSI* indicator in the oversold area (below 30); (M.vi) the steepness of the downward slope of the *SMA*; (M.vii) the *RSI* reaching a zero slope in the oversold area; (M.viii) the $D_{st}$ curve moving away from the lower Bollinger band and the start of a gradual reduction in the width of Bollinger bands, as well as a local maximum at the upper Bollinger band; (M.ix) the $D_{st}$ curve moving towards the *SMA* curve; (M.x) the exit of the *RSI* indicator from the oversold situation; (M.xi) the upward crossing of the *SMA* curve by the $D_{st}$ curve; (M.xii) the upward slope of the *SMA* curve; (M.xiii) the entrance of the *RSI* indicator to the overbought situation.

On the other hand, the secondary indications include: (S.i) the emergence of a "head fake"; (S.ii) a "divergence" between the $D_{st}$ values and the *RSI* values at local maxima or minima; (S.iii) the emergence of a "failure swing"; (S.iv) chart patterns such as "head and shoulders" or "triangles"; (S.v) the $D_{st}$ curve moving outside the Bollinger bands during its downward movement; (S.vi) the decline of the *RSI* indicator in oversold situation in combination to the downward crossing of the lower Bollinger band by the $D_{st}$ curve; (S.vii) the $D_{st}$ curve re-entrance within Bollinger bands when approaching to the end of its downward movement; (S.viii) a local minimum of $D_{st}$ values while being outside of the Bollinger bands followed by local minimum inside the bands.

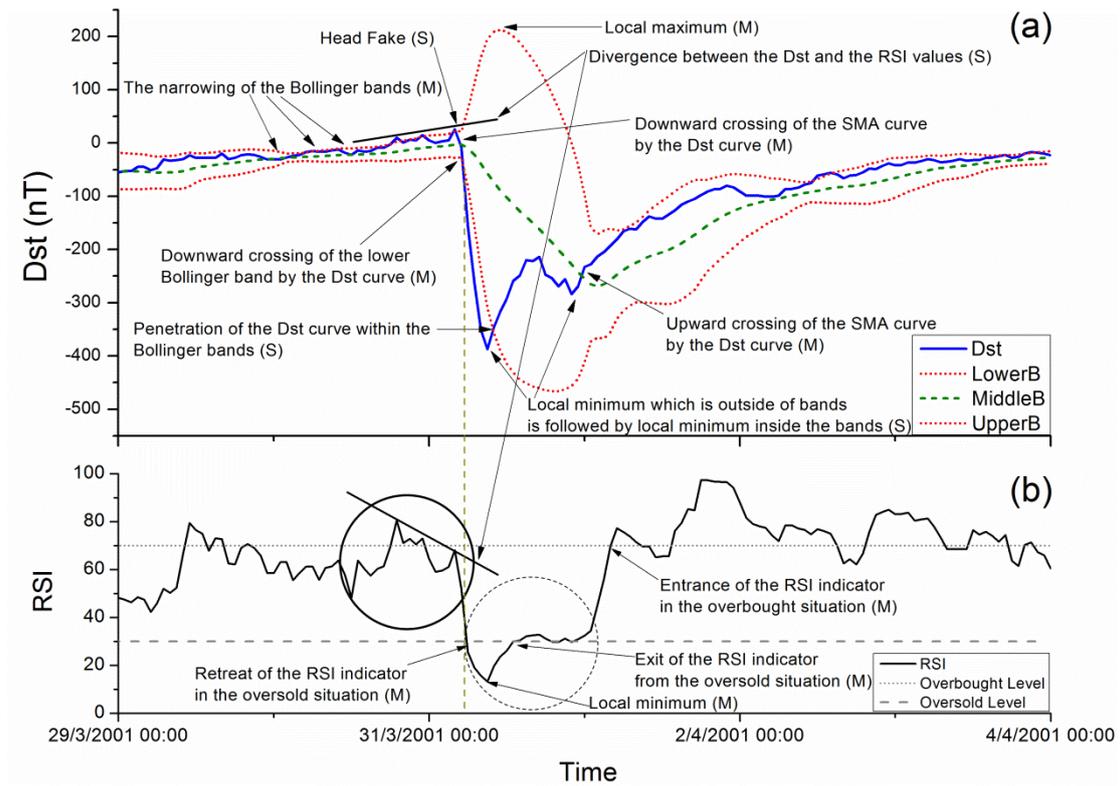

**Fig. 2.** The evolution of a magnetic storm in terms of: (a) the *SMA* and Bollinger bands, (b) the *RSI* indicator. The analyses results for a six days period from 29/03/2001 00:00 (UT) to 04/04/2001 00:00 (UT) are presented as an example. Note that the appearing "(M)" and "(S)" signs refer to the main and secondary indications, respectively, as classified in the main text of Sec. 3.2. The areas of the *RSI* curve, marked by solid and dashed circles include some secondary indications shown in detail in Fig. 3.

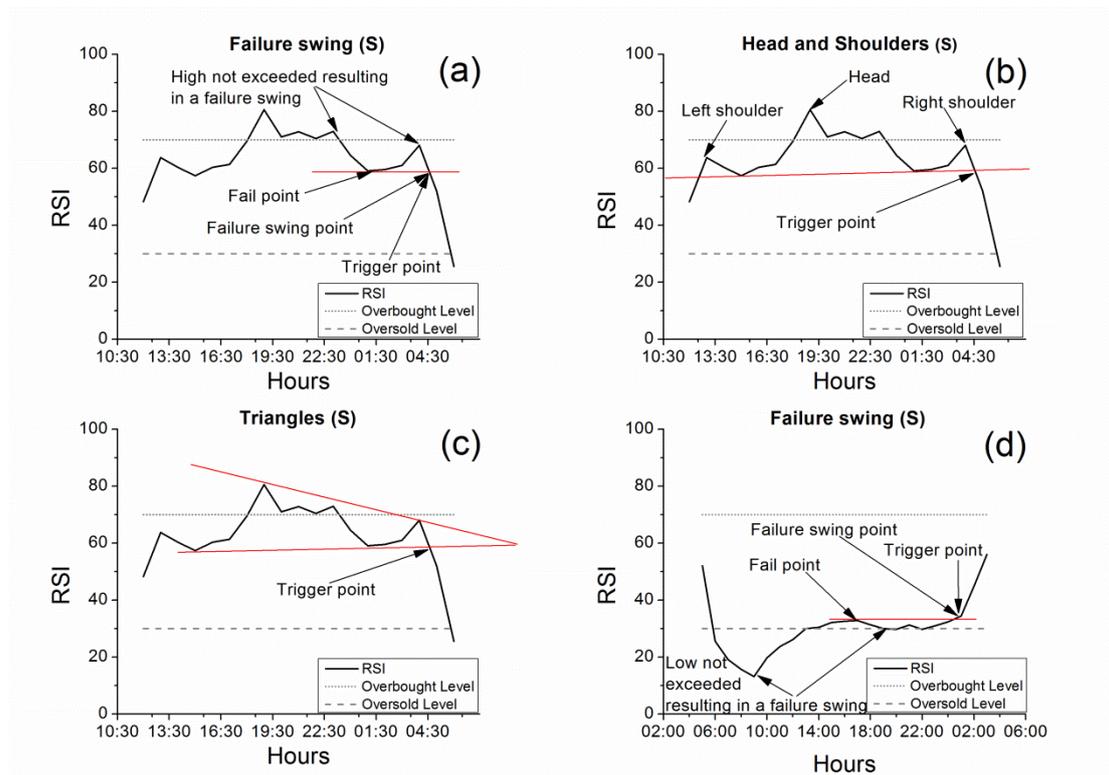

**Fig. 3.** Detail of secondary indications present in the areas of the *RSI* curve of Fig. 2.b which are marked by solid (detailed in: Fig. 3.a, b, and c) and dashed (detailed in: Fig. 3.d) circles. The examined time period lies between 30/03/2001 12:00 (UT) and 1/04/2001 03:00 (UT). Fig. 3.a and 3.d: The red line shows the failure swing point [84]. Fig. 3.b: The red line which it is called "neckline" connects the lows of the two troughs between the three peaks. The line can slope up or down. The point at which the RSI value crosses the red line is often used as a final indication for the change of trend ("trigger point") [84]. Fig. 3.c.: Horizontal support line is an imaginary line that acts as support for the RSI values. Once RSI values are below this line, it resists any upward movement of the RSI values. Down sloping top trend line is an imaginary line that acts as a layer of resistance for the RSI values [84].

In the following we present a methodology of inferring the evolution, namely the onset, duration and recovery phase, of a magnetic storm by assessing the behavior of the three considered technical analysis tools. We employ all signs, main and secondary, focusing on the temporal sequence in which they occur. The methodology is presented in the form of three consecutive phases, each one consisting of a set of indications organized in steps according to their order of occurrence. An illustrative example of application is given in parallel using a super-storm which occurred on 31/3/2001 with a peak $D_{st}$ value of -387 nT (see Fig. 2.a, as well as line no. 13 of Table 1).

*Phase I: Preparation*

(I.a) The first main indication of a possible upcoming magnetic storm is the narrowing of the Bollinger bands (cf. Fig. 2.a). The narrowing of bands indicates very low volatility of the $D_{st}$ values preceding a volatility breakout [73] and a corresponding extension of the bands. Immediately after the narrowing of the Bollinger bands we only know that this will be followed by a greater volatility of $D_{st}$ values compared to that observed during the previous period.

(I.b) In the majority of the analyzed $D_{st}$ time series cases, it has been observed that after the narrowing of the bands and before the start of the downward movement of $D_{st}$, which may evolve to a magnetic storm, some secondary evidence occur. Although such indications should only be considered as supportive of the main ones, we consider important to be mentioned for completeness purposes. One of them is a local maximum of $D_{st}$ curve which exceeds the upper Bollinger band. This local maximum often referred to as "head fake" (cf. Sec. 2.1.2) is an indication that the trend of the $D_{st}$ curve in the coming hours will be downwards (see Fig. 2.a). Also, quite often, just before the beginning of the downward movement of the $D_{st}$ curve, a "divergence" between the $D_{st}$ values and the *RSI* values appears (cf. Sec. 2.1.3, see Fig. 2.a and Fig. 2.b.), as well as a "failure swing" is observed (cf. Sec. 2.1.3, see solid cycle area in Fig. 2.b, and enlarged in Fig. 3.a), which are also indications of a continuing downward trend of the $D_{st}$ curve. Finally, during the same period there have been observed, more rarely than the previous indications, formulations in the *RSI* indicator known as "head and shoulders" or "triangles" (cf. Sec. 2.1.3, see solid cycle area in Fig. 2.b and enlarged in Fig. 3.b and Fig. 3.c, respectively).

*Phase II: Main*

(II.a) The second main indication of an approaching magnetic storm is a downward crossing of *SMA* curve by the $D_{st}$ curve which marks the beginning of a downward trend of $D_{st}$, as shown in Fig. 2.a. This is considered a very strong indication, since it has been clearly identified before each one of the studied magnetic storms. The downward movement of $D_{st}$ usually starts a few hours earlier (cf. Fig. 2.a) because, as we already mentioned, the *SMA* provides signs of trend change with a lag. Another main indication is the steep downward slope of the *RSI* indicator (Fig. 2.b). This is indicating that $D_{st}$ curve is moving with high speed downwards, and consequently the magnetic storm is in full deployment. The next very important indication, of the indications sequence implying the main phase of a magnetic storm, is the downwards crossing of the lower Bollinger band by the $D_{st}$ curve (cf. Fig. 2.a, the moment of crossing is denoted by the vertical dashed line). This usually happens shortly after the downwards crossing of *SMA* ; however, sometimes these two indications occur simultaneously. The downwards crossing of the lower Bollinger band by the $D_{st}$ curve is considered a main indication, since it has been identified in all the analyzed magnetic storm cases.

(II.b) In the majority of them, the $D_{st}$ curve, after crossing the lower Bollinger band, continues to move outside the bands during the entire time period of its downward movement. This behavior, according to technical analysis literature, is a very strong indication for the evolution of the phenomenon, since it implies a continuation of the current (downward in our case) trend [70]. However, since this is not always observed for magnetic storms it has been classified as a secondary indication. Note that in the differentiating cases after the downwards crossing of the lower Bollinger band by the $D_{st}$ curve, the $D_{st}$ curve continues moving for a while outside the bands but soon after it penetrates into the bands; however, it continues its downward movement very close, almost in parallel, to the lower Bollinger band.

(II.c) Another main indication of the deployment of the phenomenon during its main phase is the retreat of the $RSI$ indicator in the oversold area (below 30), see Fig. 2.b. Note that this has several times been observed to occur simultaneously to the downward crossing of the lower Bollinger band by the $D_{st}$ curve mentioned in step (II.a) (observe the vertical dashed line in Fig. 2.a,b). During the same time period we observe the downward slope of the $SMA$; the steeper the slope the more intensive will be the storm.

(II.d) The indication, which usually provides information about the duration of the downward movement of the $D_{st}$ curve, and correspondingly the duration of the magnetic storm, is the moment when the $RSI$ indicator declines in the oversold situation in respect with the moment of the downward crossing of the lower Bollinger band by the $D_{st}$ curve (observe the vertical dashed line in Fig. 2.a,b). Specifically, as it is mentioned in technical analysis literature: "when price touches the lower Bollinger band, and $RSI$ is above 30, it is an indication that the trend should continue, if price touches the lower Bollinger band and $RSI$ is below 30 (possibly approaching 20), the trend may reverse itself and move upward" [67]. This means that if we observe a simultaneous crossing (of the lower Bollinger band by the $D_{st}$ curve and the 30 line by the $RSI$) the magnetic storm is expected to have a long duration, while if the $RSI$ drops below 30 after the crossing of the lower Bollinger band by the $D_{st}$ curve the storm is expected to have short duration. Although the prediction of the expected duration of the downward movement of the $D_{st}$ curve according to this criterion has been proved successful for the vast majority of the cases that have been analyzed, the corresponding indication was classified as a secondary one on the grounds that there is a small number of cases for which the duration of the magnetic storm was not correctly inferred. In the case presented in Fig. 2, the $RSI$ drops below 30 almost simultaneously to the downward crossing of the lower Bollinger band by the $D_{st}$ curve which indicates that the downwards movement of the $D_{st}$ will not last long, as indeed verified by Fig. 2.a (in our example this lasts only 3 hours).

*Phase III: Recovery*

(III.a) The first main indication which provides information about the end of the phenomenon is the *RSI* curve reaching a zero slope in the oversold area. In some cases, this indication might be found earlier than the time the $D_{st}$ curve reaches its local minimum value. If the $D_{st}$ curve, after crossing the lower Bollinger band, was continuously moving outside the bands during the entire time period of its downward movement, then the re-entrance of the $D_{st}$ curve back into the bands can be taken into consideration as a secondary indication for the end of the storm. This, along with a local minimum reached by the *RSI* curve indicate an upcoming upward move of $D_{st}$.

(III.b) The following main indication is the $D_{st}$ curve moving away from the lower Bollinger band while a gradual reduction in the width of Bollinger bands is observed, as well as a local maximum of the upper Bollinger band, indicating a gradual reduction of the volatility of $D_{st}$ values. At the same time the $D_{st}$ curve is moving towards the *SMA* curve, indicating return to quiet (normal) $D_{st}$ values.

(III.c) Next, the exit of the *RSI* indicator from the oversold situation indicates that the phenomenon is coming to its end. Then the upward crossing of the *SMA* curve by the $D_{st}$ curve and the subsequent upward slope of the *SMA* curve are main indications of the end of the magnetic storm. During the same time period two secondary indications may be observed. The first of them, which might precede the exit of the *RSI* indicator from the oversold situation, is the occurrence of two local minima of the $D_{st}$ curve, one outside the Bollinger bands followed by a second inside the bands forming a W-shaped curve. The other secondary indication, which always follows the exit of the *RSI* indicator from the oversold situation, is a failure swing in the *RSI* curve implying also the upward trend of $D_{st}$ (cf. Sec. 2.1.3, see dashed cycle area in Fig. 2.b, and enlarged in Fig. 3.d).

(III.d) The final indication informing us for the termination of the phenomenon is the entrance of the *RSI* indicator to the overbought situation, which implies that the upward trend of the $D_{st}$ curve is strong before it returns to the quiet conditions.

## 4. Analysis Results

The applicability of the methodology proposed in Sec. 3 for the analysis of magnetic storms based on stock market (technical analysis) tools is examined in detail on four characteristic cases of magnetic storms. Moreover, Table 1 shows which of the considered indications (per phase and step) were identified for each one of the 24 analyzed magnetic storms. The corresponding informative plots, which summarize the analyses results for each one of magnetic storms listed in Table 1, are also provided as on-line supplementary material.

**Table 1**. List of the magnetic storms which were analyzed using the methodology proposed in Sec. 3. Apart, from the details of each magnetic storm (date/time of occurrence and peak $D_{st}$ value), the indications (per phase and step) which were identified for each one of them are denoted. The notation used is the defined in Sec. 3 for the phases, steps and the numbering of main and secondary indications. The identified indications are denoted by an "X" mark.

| M. Storm # | Occurrence date | Peak Dst (nT) | Phase I | | | | | Phase II | | | | | | | Phase III | | | | | | | | | |
|---|---|---|---|---|---|---|---|---|---|---|---|---|---|---|---|---|---|---|---|---|---|---|---|---|
| | | | Step I.a | Step I.b | | | | Step II.a | | | Step II.b | Step II.c | | Step II.d | Step III.a | | Step III.b | | Step III.c | | | | | Step III.d |
| | | | M.i | S.i | S.ii | S.iii | S.iv | M.ii | M.iii | M.iv | S.v | M.v | M.vi | S.vi | M.vii | S.vii | M.viii | M.ix | M.x | M.xi | M.xii | S.viii | S.iii | M.xiii |
| 1 | 11/02/1958 | -426 | X | | | | | X | X | X | X | X | X | | X | X | X | X | X | X | X | | X | X |
| 2 | 08/07/1958 | -330 | X | X | | | | X | X | X | X | X | X | X | X | X | X | X | X | X | X | X | | X |
| 3 | 23/04/1959 | -128 | X | X | | | | X | X | X | X | X | X | | X | X | X | X | X | X | X | | X | X |
| 4 | 15/07/1959 | -429 | X | X | | | | X | X | X | X | X | X | X | X | X | X | X | X | X | X | | | X |
| 5 | 28/10/1961 | -272 | X | X | | | | X | X | X | X | X | X | X | X | X | X | X | X | X | X | X | X | X |
| 6 | 16/02/1962 | -78 | X | | X | | X | X | X | X | | X | X | X | X | | X | X | X | X | X | | | X |
| 7 | 10/02/1963 | -62 | X | X | X | | | X | X | X | X | X | X | | X | X | X | X | X | X | X | | | X |
| 8 | 26/05/1967 | -387 | X | X | | X | | X | X | X | X | X | X | X | X | X | X | X | X | X | X | X | | X |
| 9 | 09/11/1991 | -354 | X | X | | | | X | X | X | X | X | X | X | X | X | X | X | X | X | X | | | X |
| 10 | 22/10/1999 | -237 | X | | | | | X | X | X | X | X | X | X | X | X | X | X | X | X | X | | X | X |
| 11 | 07/04/2000 | -288 | X | | X | | | X | X | X | X | X | X | X | X | X | X | X | X | X | X | X | | X |
| 12 | 16/07/2000 | -301 | X | X | | | | X | X | X | X | X | X | X | X | X | X | X | X | X | X | | X | X |
| 13 | 31/03/2001 | -387 | X | X | X | X | X | X | X | X | X | X | X | X | X | X | X | X | X | X | X | | X | X |
| 14 | 12/04/2001 | -271 | X | X | | | | X | X | X | X | X | X | | X | X | X | X | X | X | X | | | X |
| 15 | 24/11/2001 | -221 | X | X | | | | X | X | X | X | X | X | X | X | X | X | X | X | X | X | | X | X |
| 16 | 20/11/2003 | -422 | X | | | | | X | X | X | X | X | X | X | X | X | X | X | X | X | X | | | X |
| 17 | 11/02/2004 | -93 | X | | | X | | X | X | X | X | X | X | X | X | X | X | X | X | X | X | | X | X |
| 18 | 08/11/2004 | -374 | X | | X | X | | X | X | X | X | X | X | X | X | X | X | X | X | X | X | | X | X |
| 19 | 15/05/2005 | -247 | X | X | | | | X | X | X | X | X | X | X | X | X | X | X | X | X | X | | X | X |
| 20 | 24/08/2005 | -184 | X | X | | | | X | X | X | X | X | X | X | X | X | X | X | X | X | X | X | | X |
| 21 | 31/08/2005 | -122 | X | | | | | X | X | X | X | X | X | X | X | X | X | X | X | X | X | | X | X |
| 22 | 06/08/2011 | -115 | X | X | | | | X | X | X | | X | X | X | X | | X | X | X | X | X | X | X | X |
| 23 | 25/10/2011 | -147 | X | X | | | | X | X | X | X | X | X | | X | X | X | X | X | X | X | X | X | X |
| 24 | 17/03/2015 | -223 | X | X | | | | X | X | X | | X | X | X | X | | X | X | X | X | X | | | X |

### 4.1. First Case-Study

The first case we examine is that of the storm occurred on 22/10/1999 (see line no. 10 of Table 1); the analyses results are portrayed in Fig. 4. The evolution of the specific magnetic storm in terms of the phases and corresponding steps described in Sec. 3.2 follows:

*Phase I: Preparation*

In Fig. 4, we observe a (prolonged) narrowing of Bollinger bands, from 18/10/1999 21:00 until 22/10/1999 01:00. As already mentioned in step (I.a) of Sec. 3.2 this is a first indication for a possible upcoming magnetic storm. Note that the secondary indications described in step (I.b) are not observed in the specific case.

*Phase II: Main*

Step (II.a): A few hours later, on 22/10/1999 01:00, we can see that two main indications appear simultaneously, namely the downward crossing of the *SMA* curve and the lower Bollinger band by the $D_{st}$ curve. At this point we should mention that the *RSI* indicator is also moving with a steep downward slope towards the oversold (under 30) area, indicating that the $D_{st}$ values move with high speed downwards.

Step (II.b): As long as the $D_{st}$ curve moves downward outside the Bollinger bands a continuation of the current (downward) trend is implied.

Step (II.c): One hour later, on 22/10/1999 02:00, the *RSI* indicator retreats below 30 while $D_{st}$ value is -77 nT. During the next five hours, the downward movement of $D_{st}$ continues, while in parallel, the *SMA* curve retreats with steep slope, concluding the sequence of main indications related to the deployment of the magnetic storm.

Step (II.d): At this point we should note that at the moment when the $D_{st}$ curve downward crossed the lower Bollinger, the *RSI* indicator had not yet reached its lower threshold value of 30. This, as it has been observed in the majority of the examined magnetic storm cases (cf., Sec. 3.2), means that the downward movement of the $D_{st}$, and consequently the magnetic storm, is expected to last for long time, provided that the *RSI* indicator finally retreats below 30, as it indeed happens in our case.

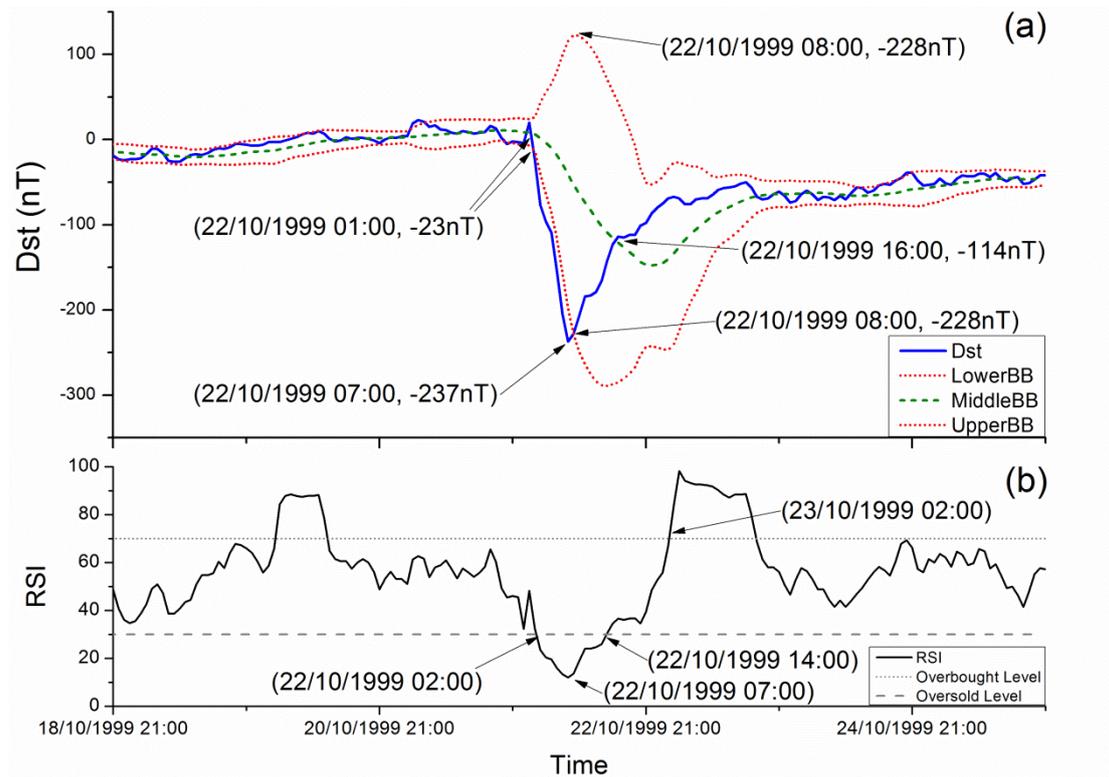

**Fig. 4.** The evolution of a magnetic storm which occurred in 1999 in terms of: (a) the *SMA* and Bollinger bands, (b) the *RSI* indicator. The analyses results for a seven days period from 18/10/1999 21:00 (UT) to 25/10/1999 21:00 (UT) are presented. The exact time and $D_{st}$ value for which specific indications were found are marked on the figure using arrows.

*Phase III: Recovery*

Step (III.a): On 22/10/1999 07:00, the *RSI* curve has almost zero slope indicating that the downward trend of $D_{st}$ curve will soon be reversed to upward trend.

Step (III.b): On 22/10/1999 08:00, while $D_{st}$ curve starts the upward movement, it enters again into the Bollinger bands, while at the same time the upper Bollinger band presents a local maximum, enhancing the indications suggesting the definite completion of the downward movement of the $D_{st}$ curve. Subsequently, we observe that the $D_{st}$ curve, as it moves upwards, moves towards the *SMA* curve, while at the same time moves away from the lower Bollinger band. Note that, in parallel the width of Bollinger bands is gradually reduced, indicating a gradual reduction of the volatility of $D_{st}$ values.

Step (III.c): After that, we identify two more main indications implying the end of the phenomenon: first the *RSI* indicator exits the oversold situation on 22/10/1999 14:00, while two hours later the upward crossing of the *SMA* curve by the $D_{st}$ curve happens. Note that the secondary indications of step (III.c) described in Sec 3.2 were not observed in the specific magnetic storm case.

Step (III.d): Finally, on 23/10/1999 02:00 the *RSI* indicator enters the overbought situation signifying the conclusion of the magnetic storm.

### 4.2. Second Case-Study

Next we examine a storm which occurred on 08/11/2004 (see line no. 18 of Table 1), while the corresponding analyses are shown in Fig. 5.

*Phase I: Preparation*

Step (I.a): First, we observe that there is a (prolonged) narrowing of the Bollinger bands from 3/11/2004 20:00 until 06/11/2004 19:00.

Step (I.b): At 21:00 of the next day we can see two secondary indications of the preparation phase, namely the head fake and the divergence between the $D_{st}$ curve and the *RSI* indicator. Moreover, in parallel, the *RSI* indicator curve forms a failure swing. Although these are secondary indications, they enhance the evidence in favor of an upcoming magnetic storm, since they imply a continuing downward movement of $D_{st}$.

*Phase II: Main*

Step (II.a): Right after, the downward crossing of the *SMA* curve by the $D_{st}$ curve is observed, while one hour later the downward movement of $D_{st}$ curve continues and crosses the lower Bollinger band. Moreover, the *RSI* indicator is moving with a steep downward slope towards the oversold (under 30) area, indicating that the $D_{st}$ values move with high speed downwards.

Step (II.b): At this point, we observe that the $D_{st}$ curve moves downward outside the Bollinger bands. As long as this is satisfied a continuation of the current (downward) trend is implied.

Step (II.c): On 08/11/2004 01:00, the *RSI* indicator curve retreats below 30. During the next six hours the downward movement of $D_{st}$ continues, while in parallel, the *SMA* curve retreats with steep slope.

Step (II.d): At this point we should take into account that the *RSI* indicator reached the low threshold value of 30 later than the moment when the $D_{st}$ curve crossed downwards the lower Bollinger. In other words, the *RSI* indicator had a value higher than 30 when the $D_{st}$ curve crossed downwards the lower Bollinger, indicating that the downward movement of the $D_{st}$ is expected to last for long time.

*Phase III: Recovery*

Step (III.a): On 08/11/2004 07:00, we observe three indications informing us about the upcoming change of the $D_{st}$ curve move to an upwards one. Namely, the *RSI* curve slope almost reaches zero, while at the same time the $D_{st}$ curve reaches its local minimum value and penetrates into the Bollinger bands.

Step (III.b) At the same time, a local maximum is observed in the upper Bollinger band, indicating that the downwards trend of $D_{st}$ curve will soon be changed to an upwards one. Indeed, one hour later the $D_{st}$ curve trend changes to upwards. From this point on the $D_{st}$ curve moves upwards approaching the *SMA* curve, departing from the lower Bollinger band, while the bands gradually narrow.

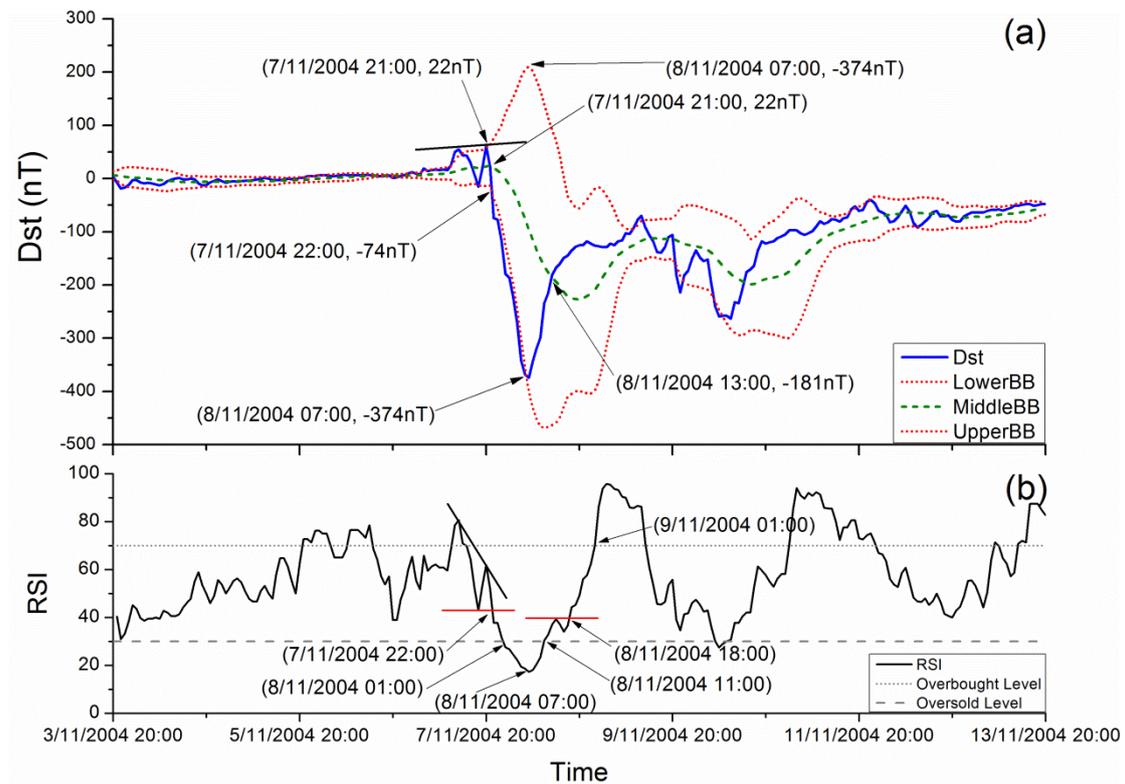

**Fig. 5**. The evolution of a magnetic storm which occurred in 2004 in terms of: (a) the $SMA$ and Bollinger bands, (b) the $RSI$ indicator. The analyses results for a ten days period from 03/11/2004 20:00 (UT) to 13/11/2004 20:00 (UT) are presented. The exact time and $D_{st}$ value for which specific indications were found are marked on the figure using arrows.

Step (III.c): Then, the $RSI$ indicator gradually moves upwards leading to an indication signifying the end of the magnetic storm: the exit of the $RSI$ indicator from the oversold situation which occurs on 08/11/2004 11:00. Following that, the next main indication is the upward crossing of the $SMA$ curve by the $D_{st}$ curve which takes place on 08/11/2004 13:00. From this point on, although with a certain delay, the $SMA$ curve develops an upwards slope. Five hours later, on 08/11/2004 18:00 a secondary indication, a failure swing, appears in the $RSI$ indicator curve enhancing the evidence in favor of the upward trend of $D_{st}$ curve.

Step (III.d): Finally, the information about the upcoming definite conclusion of the magnetic storm is provided by the entrance of the $RSI$ indicator into the overbought situation on 09/11/2004 01:00.

### 4.3. Third Case-Study

The third magnetic storm analyzed in terms of technical analysis tools is a storm which occurred on 26/05/1967(see line no. 8 of Table 1), while the corresponding analyses are shown in Fig. 6. For the specific magnetic storm we can identify the following sequence of indications which can be used to infer its evolution by short-term analysis.

*Phase I: Preparation*

Step (I.a): From 22/5/1967 00:00 up to 25/5/1967 12:00, we observe a prolonged narrowing of the Bollinger bands, the first main indication for an upcoming magnetic storm.

Step (I.b): On 25/5/1967 14:00 we observe that a head fake is formed over the upper Bollinger band, which is a secondary indication warning for an upcoming downward trend of $D_{st}$ curve. Note that this is the only secondary indication that was identified at this step for the specific storm.

*Phase II: Main*

Step (II.a): The expected downward trend of the $D_{st}$ curve indeed happened one hour later on 25/5/1967 15:00, when the $D_{st}$ curve simultaneously crossed downwards both the *SMA* curve and the lower Bollinger band. Bothe these crossings are main indications for the beginning of a magnetic storm. Turning now to the *RSI* indicator we can see that although it was moving with a steep downward slope towards the oversold situation, indicating a high downward speed for $D_{st}$ values, it finds support at the 30 level. During the next five hours, the $D_{st}$ curve moves inside the Bollinger bands, close to the lower band. However, the width of the bands continues to expand. On 25/5/1967 21:00 we observe that $D_{st}$ curve crosses once more the lower Bollinger band downwards. Note that between these two downward crossings of the lower Bollinger band the $D_{st}$ curve moves inside the bands, i.e. step (II.b) has not yet been observed. As already mentioned in Sec. 3.2, this step corresponds to a secondary indication which is not always observed after a downwards crossing of the lower Bollinger band by the $D_{st}$ curve, so we don't pay any special attention to this fact. After, the second crossing of the lower Bollinger band, we also notice that d the *RSI* indicator moves with a steep downward slope, which is gradually reducing.

Step (II.b): From 25/5/1967 21:00 onwards, the $D_{st}$ curve remains outside the Bollinger bands, indicating a continuation of its current (downwards) trend.

Step (II.c): The *RSI* indicator moves again downwards to the oversold situation to retreat below the 30 level one hour later. At the same time we note that the *SMA* curve gradually begins to move downwards with steep slope, indicating a strong downwards trend for $D_{st}$.

Step (II.d): Given that the $D_{st}$ curve downward crossed the lower Bollinger band before the *RSI* indicator retreats below 30, we expect that the downward movement of $D_{st}$ until it reaches its lowest value will last for long time.

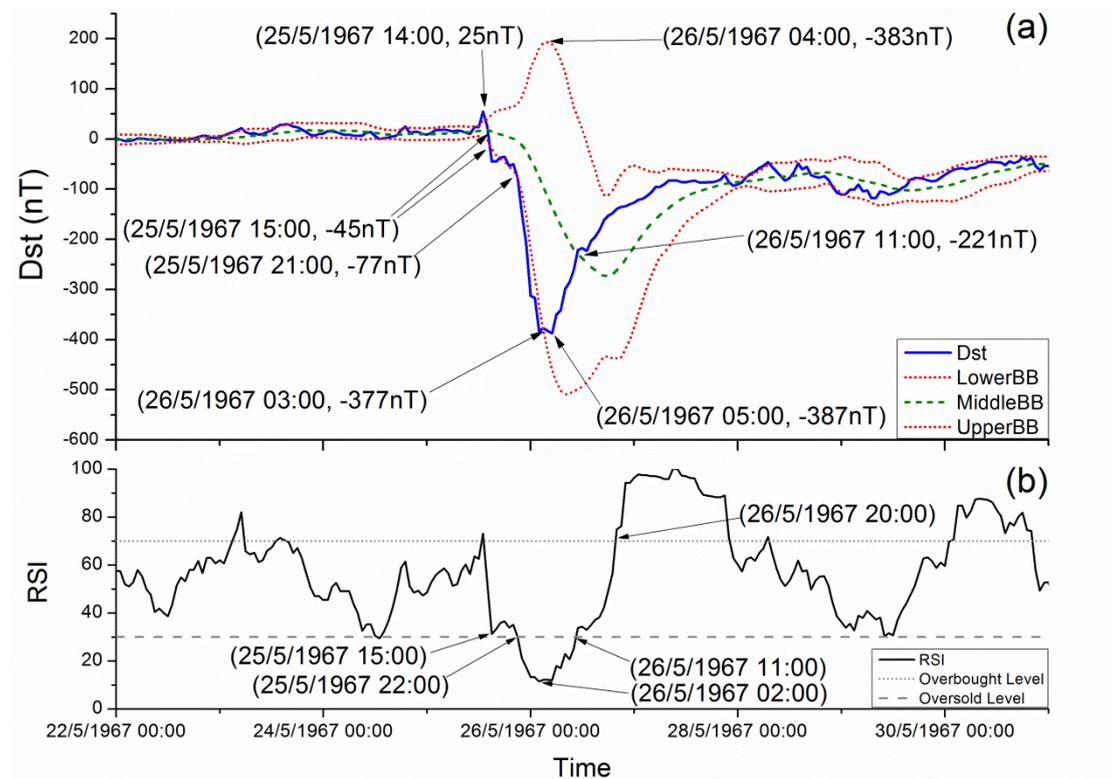

**Fig. 6.** The evolution of a magnetic storm which occurred in 1967 in terms of: (a) the *SMA* and Bollinger bands, (b) the *RSI* indicator. The analyses results for a nine days period from 22/05/1967 00:00 (UT) to 31/05/1967 00:00 (UT) are presented. The exact time and $D_{st}$ value for which specific indications were found are marked on the figure using arrows.

*Phase III: Recovery*

Step (III.a): On 26/5/1967 02:00 the *RSI* curve reaches almost zero slope indicating that the downward trend of $D_{st}$ curve is soon expected to be reversed to an upward trend. One hour later, at 03:00, we can see that the $D_{st}$ curve penetrates inside the Bollinger bands, which is a secondary indication enhancing the possibility that the $D_{st}$ trend is soon expected to be upward.

Step (III.b): One hour later, on 25/5/1967 04:00 a local maximum of the upper Bollinger band is observed which is a main indication also informing us about the upcoming upwards trend of the $D_{st}$ curve. Next, the $D_{st}$ curve moves towards the *SMA* curve and away from the lower Bollinger band, while, in parallel, the width of the bands is gradually reducing.

Step (III.c): On 26/5/1967 11:00 two main indications appear simultaneously indicating the end of the phenomenon. Specifically, the $D_{st}$ curve upward crosses the *SMA* curve, and soon after the *SMA* develops an upwards slope, while the *RSI* indicator exits the oversold situation.

Step (III.d): Finally, on 26/5/1967 20:00 we observe the *RSI* indicator entering the overbought situation which signifies the definite completion of the phenomenon.

### 4.4. Fourth Case-Study

The last magnetic storm case is a relatively recent intense storm which occurred on 25/10/2011 (see line no. 23 of Table 1); the corresponding analyses are shown in Fig. 7. The evolution of the specific magnetic storm in terms of the phases and corresponding steps described in Sec. 3.2 follows:

*Phase I: Preparation*

Step (I.a): From 20/10/2011 21:00 up to 24/10/2011 19:00 we observe a prolonged narrowing of the Bollinger bands, which, as already mentioned in Sec. 3.2, is a main indication for the preparation of an upcoming magnetic storm.

Step (I.b): On 24/10/2011 20:00, while until then the $D_{st}$ curve was moving mainly within Bollinger bands, we observe a head fake which foreshadows an upcoming downtrend curve $D_{st}$. Other secondary indications, of the ones corresponding to the specific step in Sec. 3.2, were not observed for the specific magnetic storm.

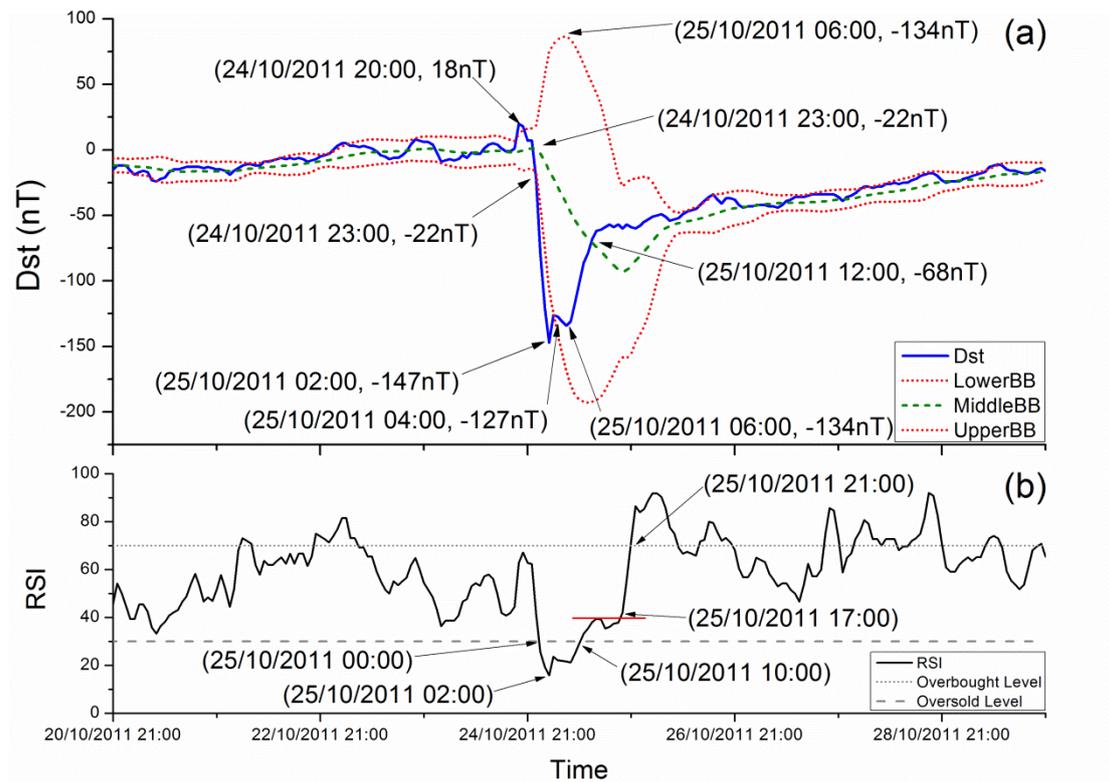

**Fig. 7.** The evolution of a magnetic storm which occurred in 2011 in terms of: (a) the *SMA* and Bollinger bands, (b) the *RSI* indicator. The analyses results for a nine days period from 20/10/2011 21:00 (UT) to 29/10/2011 21:00 (UT) are presented. The exact time and $D_{st}$ value for which specific indications were found are marked on the figure using arrows.

*Phase II: Main*

Step (II.a): A few hours later, on 24/10/2011 23:00, we can see that two main indications appearing simultaneously, namely the downward crossing of the *SMA* curve and the lower Bollinger band by the $D_{st}$. We also observe that the *RSI* indicator is moving with a steep downward slope towards the oversold (under 30) area, indicating that the $D_{st}$ values move with high speed downwards.

Step (II.b): After the crossing of the lower Bollinger band by the $D_{st}$ curve, the $D_{st}$ curve moves downward outside the Bollinger bands signifying a continuation of the current (downward) trend.

Step (II.c): On 24/10/2011 00:00 the *RSI* indicator enters the oversold situation, while from this point on the *SMA* curve moves downwards with a steep slope.

Step (II.d): The specific magnetic storm is an example for which step (II.d) does not provide a correct prediction for the duration of the magnetic storm. Although the retreat of the *RSI* indicator into the oversold situation happened delayed with respect to the moment when the downward crossing of the lower Bollinger band by the $D_{st}$ curve occurred, the duration of the downward movement of the $D_{st}$ curve, and correspondingly the duration of the magnetic storm, was not long. We remind that the specific indication was classified as a secondary one exactly due to the reason that there are magnetic storm cases, even though very few, for which it doesn't lead to a correct inference for their expected duration.

*Phase III: Recovery*

Step (III.a): On 25/10/2011 02:00, we observe that the *RSI* indicator almost reaches zero slope, indicating that $D_{st}$ curve will possibly move upwards during the following hours. The $D_{st}$ curve also reaches its local minimum value then, while two hours later the $D_{st}$ curve enters back into the Bollinger bands, one more indication that the phenomenon is leaded to its completion.

Step (III.b): Following that, on 25/10/2011 06:00, we observe a local maximum of the upper Bollinger band signifying the upcoming reduction of $D_{st}$ values' volatility, as well as a secondary indication, usually observed later and for this reason described as part of step (III.c), the occurrence of two local minima of the $D_{st}$ curve, one outside the Bollinger bands followed by a second inside the bands forming a W-shaped curve. During the next hours, the $D_{st}$ curve develops an upwards slope towards the *SMA* curve, while at the same time width of Bollinger bands is gradually reduced and the $D_{st}$ curve moves away from the lower Bollinger band, all implying a return to normal (quiet) $D_{st}$ values.

Step (III.c) On 25/10/2011 10:00 the *RSI* curve exits the oversold situation, while two hours later the $D_{st}$ curve upward crosses of the *SMA* curve, which then moves with an upwards slope, enhancing the indications that the phenomenon is coming to its end. Moreover, on 25/10/2011 17:00 a failure swing is observed, which is a secondary indication.

Step (III.d): The final recovery phase indication, signifying the conclusion of the magnetic storm, comes on 25/10/2011 21:00 when the *RSI* indicator enters the overbought situation.

## 5. Discussion - Conclusions

In the frame of complex systems, we studied the time series of $D_{st}$ .in terms of the empirical financial analysis method known as technical analysis, focusing on the temporal evolution of magnetic storms. Specifically, we employed the combination of three very popular tools of technical analysis, the simple moving average ($SMA$), the Bollinger bands, and the relative strength index ($RSI$) in order to formulate a methodology of magnetic storm analysis which could be used for space weather forecasting.

This methodology was developed after the analysis of more than 20 cases of magnetic storms, revealing all indications which, in specific temporal sequence of occurrence, provide information about the onset, duration and recovery phase of a magnetic storm. The applicability of the proposed methodology was presented in detail on four characteristic cases of magnetic storms, while the results for the whole set of 24 analyzed magnetic storms verify the repeatability of the proposed indications, rendering the results encouraging for space weather forecasting. However, we focus on the fact that the results of this study enhance the view that quantitative analysis methods can be successfully be transferred between economic and geophysical systems. Our results show that $D_{st}$ time series around the occurrence of magnetic storms can be successfully analyzed by the same empirical tools applied on share price time series for investment analysis.

In general, when a new magnetic storm is about to happen, the sequence of indications described by the proposed methodology are repeated. However, in some special cases such as "double storms" (two magnetic storms occurring close to each other), very intensive, or very long storms, some indications may not be observed, implying the need for further investigation of the proposed use of the considered technical analysis tools. This should mainly focus on the tuning of analysis parameters based on extensive statistics resulting from application of the specific tools for long enough time periods, e.g., for the last two solar cycles. Note that the considered technical analysis methods were applied on $D_{st}$ time series directly adopting, mutatis mutandis in terms of sampling rate scale, the parameter values usually employed for the analyses of financial time series. Specifically, the usually employed parameter values for the calculation of $SMA$ and Bollinger bands when applied on daily stock market share data $(n = 20, K = 2)$, were also used for the case of hourly $D_{st}$ data.

The proposed method, after an appropriate tuning of the analysis parameters, could be combined with other statistical time series analysis methods, like different kind of entropies, Hurst exponent, fractal analysis, and log-periodic corrections in order to lead to an automated magnetic storm forecasting tool with an acceptable success rate (based on extensive statistics), at least for the classes of intense and super storms. Although this is outside the scope of the specific work, it is in our near future plans to elaborate such a study.

The here presented study followed the empirical way of analyzing financial time series for the analysis of $D_{st}$ time series, trying to identify indications, also used in economics, which could be associated with the temporal evolution of a magnetic storm. In a next step, one could focus on the interpretation of the observed behavior of the considered financial tools, in terms of the physics of the magnetosphere. Such a study could, on one hand, increase the effectiveness of the proposed methodology for the study of magnetic storms, and, on the other hand, offer ideas in order to better understand stock market processes.